\documentclass[12pt]{iopart}

\usepackage[utf8]{inputenc}
\usepackage[T1]{fontenc}
\usepackage{graphicx}
\usepackage{dcolumn}
\usepackage{bm}
\usepackage{epsfig}
\usepackage{subfigure}
\usepackage{graphics}
\usepackage{epstopdf}
\usepackage[english]{babel}
\usepackage{amsfonts}
\usepackage{mathrsfs}
\usepackage{amsthm}
\usepackage{amssymb}
\usepackage{enumerate}
\usepackage{subfigure}
\usepackage{algorithm}
\usepackage{algorithmicx}
\usepackage{algpseudocode}
\usepackage{hyperref}
\usepackage{multirow}
\floatname{algorithm}{Algorithm}
\usepackage{iopams}
\begin{document}

\title{Graphical Representation and Hierarchical Decomposition Mechanism for Vertex-Cover Solution Space}

\author{Wei Wei$^{1,2,4,5}$, Xiangnan Feng$^{6,*}$, Jiannan Wang$^{2,3}$, Xue Liu$^{1,2,5}$, Zhiming Zheng$^{1,2,3,4,5}$}

\address{1. School of Mathematical Sciences, Beihang University, Beijing, 100191, People's Republic of China\\
2. Key Laboratory of Mathematics, Informatics and Behavioral Semantics, Ministry of Education, 100191, People's Republic of China\\
3. Research Institute of Frontier Science, Beihang University, Beijing, 100191, People's Republic of China\\
4. Beijing Advanced Innovation Center for Big Data and Brain Computing, Beihang University, Beijing, 100191, People's Republic of China\\
5. Peng Cheng Laboratory, Shenzhen, Guangdong, 518055, People's Republic of China\\
6. Center for Humans and Machines, Max Planck Institute for Human Development, Berlin, 14195, Germany}

\ead{fengxiangnan@buaa.edu.cn}
\vspace{10pt}
\begin{indented}
\item[]September 2021
\end{indented}

\begin{abstract}
In this paper, solution space organization of minimum vertex-cover problem is deeply investigated using the K\"{o}nig-Eg\'{e}rvary (KE) graph and theorem, in which a hierarchical decomposition mechanism named KE-layer structure of general graphs is proposed to reveal the complexity of vertex-cover.
An algorithm to verify the KE graph is given by the solution space expression of vertex-cover,
and the relation between multi-layer KE graphs and maximal matching is illustrated and proved.
Furthermore, a framework to calculate the KE-layer number and approximate the minimal vertex-cover is provided,
with different strategies of switching nodes and counting energy.
The phase transition phenomenon between different KE-layers are studied with the transition points located,
and searching of vertex-cover got by this strategy presents comparable advantage against several other methods. 
The graphical representation and hierarchical decomposition provide a new perspective to illustrate the intrinsic complexity for large-scale graphs/systems recognition.
\\
\\
{\bf keywords.}
KE-Layer Structure, Minimum Vertex Cover, Random Graphs, Solution Space Expression

\end{abstract}
%
%
%
%
%

\section{Introduction}
As the development of the computer science and artificial intelligence, the concern on data analysis and objective optimization becomes increasingly important and many cutting-edge technologies in machine learning and deep learning greatly promote the current research in a variety of fields \cite{friedman2001elements, goodfellow2016deep}. Core difficulty faced by these technologies is commonly and mainly the high computational complexity induced by the real-world applications, and different approximating strategies on the computing-hard problems (i.e., NP even NP-complete problems) have achieved a great success in game theory, natural language processing, image processing etc., assisted with the computer software and hardware technologies. For example, exhaustive searching in game theory can be substituted by strategic computation using deep reinforcement learning \cite{silver2017mastering}, feature extraction and pattern recognition in image processing can be automatically accomplished by convolution and pooling in convolution neural network \cite{1995Artificial}, and graph partition in combinatorial optimization can be calculated by node coding using graph representation learning \cite{2019Approximation}. A large range of NP and NP-complete problems in various applications provide scenarios for the nowadays artificial intelligence technology.

Combined with tools from statistics and graph theory, a large number of researches from graph-based complex systems (networks) \cite{newman2010networks} have been applied in many engineering and natural science fields, like biology \cite{zitnik2018modeling}, social structure \cite{centola2018experimental}, world wide web (WWW) \cite{6621057} and human dynamics \cite{alessandretti2018evidence}. To transform irregular complex systems into regularized organizations/representations is always an active topic in system learning, e.g., the images are treated as tensor in convolution neural network \cite{2015Deep}, and the graph nodes can be expressed as vectors in graph representation learning \cite{2016Semi, graphattention, 2020Graph}. For NP or NP-complete problems such as combinatorial optimization problems based on graphs, expressing underlying roles or relations of the nodes is also a kernel idea to solve and understand their complexity. In statistical mechanics, the nodes on a graph of combinatorial optimization problems are assigned marginal probabilities in the solution space by the replica and cavity method \cite{mezard2003alternative, Mertens2005Threshold}, which can also be viewed as nodes' representation to reveal the underlying values and relations. The representation mechanism of network/graph aims at decoupling the complicated relations among the nodes to simplify the whole system, which is generally focused on in the nowadays representation learning research \cite{Forsyth2015Representation, 2017Representation}. In the classical linear algebraic system, its solution spaces can be represented by a linear space composed of some basis vectors \cite{lang1984alge}, and in a class of pure nonlinear algebraic system there is still a group of basis representation for all of its solutions \cite{wei2010mas}. Meanwhile, hierarchical decomposition of networks/graph provides effective understanding of the complicated system structure and functions: centrality measurements and community detecting methods like \emph{k}-shell \cite{kitsak2010identification}, \emph{k}-core \cite{dorogovtsev2006kcore} and modularity \cite{newman2006modularity}, and many leaf-removing strategies \cite{bollobas1985random, mezard2003alternative, weigt2001physe, Mertens2005Threshold, wei2009jstat} are commonly used, which could illustrate the layer structure of graphs and point out the hubs.

Existing representation studies on the NP problems mainly concern on the node status representation, and we have proposed a strict graphical representation of the whole solution space of minimum vertex-cover problem when the graph topology is bipartite \cite{wei2015birsg}, which combined the node statuses with the edge statuses. In this paper, the corresponding results for minimum vertex-cover problem will be strictly generalized on the famous K\"{o}nig-Eg\'{e}rvary (KE) graphs, and combined with the minimum coverage requirements, a hierarchical decomposition mechanism based on the solution space expression of the KE-subgraphs will be deeply investigated to reveal the intrinsic computational hardness of the NP-complete problem. The further research on graphical representation and hierarchical decomposition mechanism for vertex-cover solution space gives a representative instance for understanding the complexity of NP-complete problems, and could provide insightful recognization and efficient way in analyzing graph-based complex systems.

\section{Preliminary of KE Graph and Solution Space Expression for Vertex-Cover}

In this section, some basic knowledge about the K\"{o}nig-Eg\'{e}rvary theorem with KE graphs and solution space expression technique for vertex-cover will be introduced as preliminary.

\subsection{Relations of KE Graph with Minimum Vertex-Cover}

For a graph $G$ composed by the vertex set $V$ with $n$ vertexes and edge set $E$ with $m$ edges, a vertex-cover is a vertex subset $C(G)\subset V$ that for any edge $(v_i,v_j)\in E$ there exists $v_i\in C(G)$ or $v_j\in C(G)$. If the cardinality of vertex-cover set achieves the lowest, it is called the \emph{minimum vertex-cover} with the corresponding cardinality named as \emph{minimum vertex-cover number} of a graph, and for a graph there could exist more than one minimum vertex-covers \cite{biggs1986graph}. The problem of finding the minimum vertex-cover belongs to Karp's 21 NP-complete problems \cite{Karp1972ibm} and the 6 basic NP-Complete problems \cite{Cook1971acm}, which have broad applications in real world but have been recognized to be intrinsically hard to solve \cite{Papadimitriou2003computational}.

Another widely studied topic in graph theory is the maximum edge matching \cite{biggs1986graph}. An edge matching, or independent edge set, of a graph is an edge subset $M\subset E$ satisfying that any two edges in $M$ have no common vertex, and a vertex is matched if it is an endpoint of one edge in $M$. The edge matching set with the largest cardinality is called the \emph{maximum (edge) matching} with  the corresponding cardinality named as \emph{maximum matching number} of a graph, and there could exist more than one maximum matchings for a graph.
Finding the maximum matching for a general graph belongs to P problems,
namely for a graph its maximum matching could be found in polynomial time.
The blossom algorithm is widely applied to find the maximum matching for general graphs \cite{edmonds1965paths},
while the Hungarian maximum matching algorithm is widely used to find the maximum matching for bipartite graphs \cite{kuhn1956variants}.
The maximum matching has a large number of applications, e.g.,
this concept is deeply related to Kekulization and the process of Tautomerization in chemistry \cite{may2015cheminformatics}.

Bipartite graph can associate the minimum vertex-cover with its maximum matching.
A graph $G$ is a \emph{bipartite} one if there exist two non-empty node subsets $V_1$ and $V_2$ with $V_1\cap V_2 = \emptyset$ and $V_1\cup V_2 = V$, satisfying that for any edge $(v_i,v_j)\in E$, $v_i$ and $v_j$ must belong to different subsets $V_1$ or $V_2$.
The bipartite graph could be used to illustrate the relation between objects with two different types especially in social network study \cite{benchettara2010supervised},
e.g., the relations between actors and movies.
One of the most crucial property of bipartite graph, given by K\"{o}nig-Eg\'{e}rvary theorem, is that its maximum matching number is equal to its minimum vertex-cover number. This property builds a bridge between the maximum matching problem and minimum vertex-cover problem and makes it possible to find the minimum vertex-cover of bipartite graphs in polynomial time \cite{storer2012introduction}.

A graph whose maximum matching number equals to its minimum vertex cover number is called the K\"{o}nig-Eg\'{e}rvary (KE) graph.
There exist more types of KE graphs other than the bipartite ones, and a commonly studied one is the No-Leaf-Removal-Core graph.
In minimum vertex-cover problem, a leaf is regarded as a one-degree node and its neighbor with connecting edges \cite{weigt2001physe},
which suggests exactly one cover number and at most one edge matching in a leaf.
If there exist only isolated nodes or no node after all the leaves are removed iteratively,
this graph is called \emph{No-Leaf-Removal-Core graph} \cite{feng2019core}.
For the No-Leaf-Removal-Core graph, its minimum vertex-cover number equals to its leaf number, and since different leaves are independent subgraphs (no common nodes and edges), its maximum matching number should be larger than its leaf number;
for the vertex cover of a graph, each independent edge should occupy at least
one cover number, and the minimum vertex-cover number must be larger than its maximum matching number.
Thus, the No-Leaf-Removal-Core graph should
have its maximum matching number equal to its minimum vertex-cover number, and it is one type of KE graph.

\subsection{Solution Space Expression of Vertex-Cover}

For the minimum vertex-cover problem, our aim is to locate the least number of covered nodes and make each edge on the given graph touched by
at least one such node. Generally, one given graph owns many minimum vertex-covers which form a solution space for this combinatorial optimization problem.

From the former work \cite{wei2012rsg}, it has been already known that on structures of KE graphs including trees, no leaf-removal-core graphs and bipartite graphs, for minimum vertex-cover problem, we have corresponding algorithms to achieve the solution space expression, which is called \emph{reduced solution graph}. The reduced solution graph provides a complete description of all the minimum vertex-cover solutions, and it can be easily obtained on these topologies \cite{wei2015birsg}. On the reduced solution graph expression, all the nodes are classified as \emph{uncovered backbones}, \emph{covered backbones} and \emph{free nodes}; nodes should always be uncovered or covered separately in all the solutions, namely \emph{frozen} nodes, or
could have alternative assignments on different solutions, namely \emph{unfrozen} nodes; all edges are classified as single or double edges, in which double edges suggest the \emph{mutual-determination} relations between two unfrozen nodes and these two nodes will mutually affect the values of each other (two free nodes in a mutual-determination relation must have one and only one covered node) and the single edges still follow the coverage requirement.

\begin{figure}[htbp]
	\centering
	\includegraphics[scale=0.64]{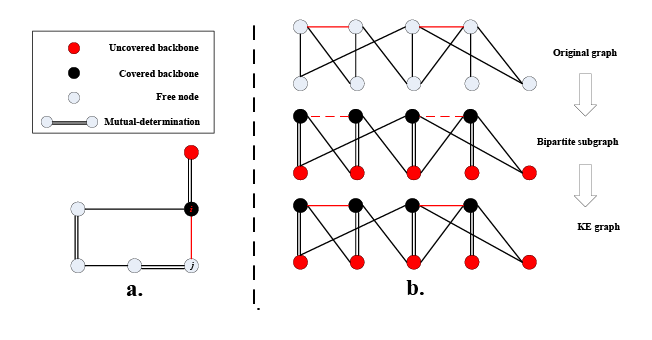}	
	\caption{A schematic view of the graphical representation and hierarchical decomposition mechanism for vertex-cover solution space. Only double edges between free nodes on the reduced solution graph represent mutual-determinations, and others correspond to a group of maximum matching in the absence of ambiguity. \textbf{a} shows an example of exact solution space expression; \textbf{b} illustrates how to achieve a solution space expression for a KE graph.}\label{schematic}
\end{figure}

Using the representation of the three classes of nodes and the two types of edges on the original graph, the vertex-cover solution space could be sufficiently expressed for a large class of graph instances, and indeed most of the known easily-solving minimum vertex-cover instances belong to KE graphs \cite{wei2012rsg,wei2015birsg}, which have exact solution space expressed by the node and edge class labels.
In Figure \ref{schematic}, a schematic view of the solution space expression is provided. \textbf{a} shows the different classes of nodes and edges, and gives an instance for minimum vertex-cover solution space of a simple graph. In this graph, node $i$, marked by black, has to be covered in all solutions, which makes its one-degree neighbor node, marked by red, namely uncovered all the time; the double edges connect the free nodes, and since on each double edge, one node getting covered will make the other one uncovered, the two double edges in the graph lead to two covered nodes. Thus in this case, the minimum vertex-cover number of the graph is 3, and all possible solutions are presented by the marked nodes and different type of edges. \textbf{b} is an example of getting a solution space expression for a KE graph, of which detailed procedures are introduced in \cite{wei2015birsg}.


\section{Solution Space Expression Algorithm for Vertex-Cover of KE Graph}
In this section, an algorithm based on the solution space expression will be provided to judge whether a graph is a KE graph. Meanwhile, the vertex-cover solution space expression for KE graph can be obtained accordingly.

Since the edges in the maximum matching are independent and all the covered nodes must be matched nodes in a KE graph, there must be one and only one covered node in each matching edge, and all the unmatched nodes must be uncovered backbones in the minimum vertex-cover solution space of KE graph. Thus, there is a natural correspondence between the concept of edge matching and the mutual-determination relation on KE graphs, which also could be used to determine whether a graph is a KE one. The algorithm to examine the KE graph and get the solution space expression of its minimum vertex-cover is represented in Algorithm \ref{main_process}, which involves the following steps:

\begin{algorithm}[htbp]
	\caption{Get Solution Space Expression for KE Graph Vertex-Cover}
	\begin{algorithmic}
	         \State {\bf Input:} Graph $G$; {\bf Output:} Reduced solution graph $S(G)$ of $G$
	         \State
	         \State $M=$ Maximum matching of $G$, Initialize $S(G)=G$;
	          \For{Any match (\emph{i,j}) in $M$}
	          \State Set $(i,j)$ a double edge in $S(G)$;
	          \EndFor
	          \While{Any unmatched node $k$ exists}
	          \State Set $k$ uncovered backbones in $S(G)$;
	          \State $S(G)$\emph{=\textbf{Freezing-Influence}}($S(G),k$);
	          \EndWhile
	          \State {$S(G)$\emph{=\textbf{Conflict-Checking}}($S(G)$)};
	          \State * $S(G)=\emptyset$ means graph $G$ is not a KE graph.
	\end{algorithmic}
	\label{main_process}
\end{algorithm}

$\textbf{Step\ 1:}$ For the graph, find one of its maximum matching, which could be done in polynomial time. If the graph is a KE one, then all the covered nodes should be matched.

$\textbf{Step\ 2:}$ Assign double edges for each matched edge and mark the unmatched nodes as uncovered backbones, do the \emph{Freezing-Influence} described in Algorithm \ref{freezing_influence} on the graph, which makes sure that all the neighbors of an uncovered backbone should be covered backbones, and the mutual-determination neighbors of a covered backbone should be an uncovered backbone. If there is some \emph{conflict} while determining the state of some nodes (i.e., some nodes are required to be covered by one neighbor and uncovered by another simultaneously), the cover number of the whole graph will exceed the maximum matching number and it cannot be a KE graph.

$\textbf{Step\ 3:}$ Apply the \emph{Conflict-Checking} from Algorithm \ref{conflict_checking} for the rest unfrozen nodes after \textbf{Step\ 2}, which involves the \emph{Freezing-Influence} and \emph{Consistency-Checking} in Algorithm \ref{freezing_influence}. Considering unfrozen node $i$, set $i$ as covered and uncovered separately and check whether there is conflict: if there are conflicts on both cases, the graph is not a KE graph and additional cover number is inevitable; if there is only conflict on one case (say the covered case), node $i$ should be kept away from this case (it should be an uncovered backbone); if there is no conflict on both cases, node $i$ should be a free node. For all the new produced backbones, the freezing influence operation should be also performed.

$\textbf{Step\ 4:}$ If the process of \textbf{Step\ 3} can survive after all the nodes are checked, it is a KE graph and its solution space expression of minimum vertex-cover could be got. Otherwise it is not a KE graph.

\begin{figure}[htbp]
	\centering
	\includegraphics[scale=0.23]{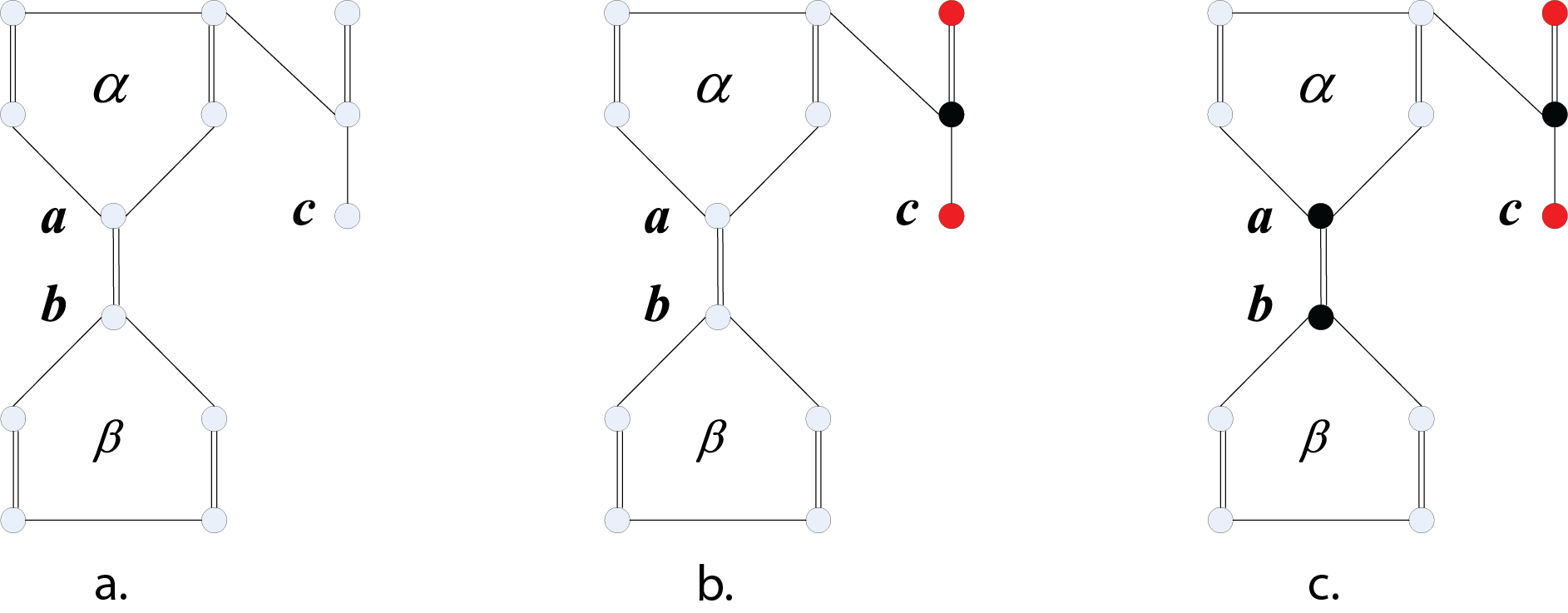}
	
	\caption{An instance for determining whether a graph is a KE one, with red nodes for uncovered backbones, black nodes for covered backbones and white nodes for free nodes. \textbf{a} is the original graph with its maximum matching from \textbf{Step 1}. \textbf{b} gives an unmatched node $c$ and performs a \emph{Freezing-Influence} operation from \textbf{Step 2}. \textbf{c} checks the nodes in the two cycles $\alpha, \beta$, and finds that nodes $a$ and $b$ must be both covered backbones by \textbf{Step 3}. But nodes $a$ and $b$ form a matching with two covered nodes, which conflicts with the KE property, and in this instance it is not a KE graph. }\label{instance}
\end{figure}

\begin{algorithm}[htbp]
	\caption{KE Graph Verification Functions}
	\begin{algorithmic}
		\Function{\textbf{\emph{Freezing-Influence}}} {$S(G),k$}
		\For{Any $k$'s neighbor $j$}
		\State Set $j$ covered backbones in $S(G)$;
		\EndFor
		\While{Any double edge ($i,j$) has node $i$ covered backbone and $j$ unfrozen}
		\State Make $j$ uncovered backbone in $S(G)$;
           	\State {$S(G)$\emph{=\textbf{Freezing-Influence}}($S(G),j$)};
		\State \textbf{Return} $S(G)$
		\EndWhile
		\EndFunction
		\State
		\Function{\textbf{\emph{Consistency-Checking}}} {$S(G)$}
		\For{Any uncovered backbone $i$ in $S(G)$}
		\If{There exist uncovered backbone neighbors of $i$}     
		\State \textbf{Return} 1 
		\EndIf
		\EndFor
                 \For{Any double edge in $S(G)$}
		\If{Its two ends are both covered backbones}     
		\State \textbf{Return} 1
		\EndIf
		\EndFor	
		\State \textbf{Return} 0	
		\EndFunction	
	\end{algorithmic}
	\label{freezing_influence}
\end{algorithm}

%
\begin{algorithm}[htbp]
	\caption{Conflict Checking Process}
	\begin{algorithmic}
		\Function{\textbf{\emph{Conflict-Checking}}} {\emph{$S(G)$}}	
		\State \textbf{\emph{Consistency-Checking}}(\emph{$S(G)$})
		\For{Any unfrozen node $i$ in $S(G)$}
		\State Set $S^{i+}(G)=S(G)$, and set $i$ uncovered backbone in $S^{i+}(G)$;
		\State $S^{i+}(G)$=\textbf{\emph{Freezing-Influence}} ($S^{i+}(G),i$);
	         \State $posi=$\textbf{\emph{Consistency-checking}} (\emph{$S^{i+}(G)$});
	         \State Set $S^{i-}(G)=S(G)$, and set $i$ covered backbone in $S^{i-}(G)$;
	         \If{$i$ has a double-edge free neighbor $j$}
	         \State Set $j$ uncovered backbone in $S^{i-}(G)$;
	         \State $S^{i-}(G)$=\textbf{\emph{Freezing-Influence}} ($S^{i-}(G),j$);
	         \State $nega=$\textbf{\emph{Consistency-Checking}} {(\emph{$S^{i-}(G)$})};
	         \EndIf
	         \If{\emph{posi}=1 and \emph{nega}=1}\State $S(G)=\emptyset$, \textbf{break};\EndIf
	         \If{\emph{posi}=1 and \emph{nega}=0}\State $S(G)=S^{i-}(G)$;\EndIf
	         \If{\emph{posi}=0 and \emph{nega}=1}\State $S(G)=S^{i+}(G)$;\EndIf
	         \State \textbf{Return} $S(G)$
	         \EndFor
	         \EndFunction
	         \end{algorithmic}
	\label{conflict_checking}
\end{algorithm}
For the above process, conflict analysis is the key to determine whether the graph is a KE one, and the conflict could come from \textbf{Step\ 2} and \textbf{Step\ 3}. For the conflict in \textbf{Step\ 2}, it results from the KE property and requirement of minimum coverage, and it is an exact step. For the conflict in \textbf{Step\ 3}, node $i$ cannot have values ensuring the KE property for both conflict cases; for the single conflict case, it indeed fulfills the \emph{odd cycle breaking} \cite{wei2012rsg}, which makes it also an exact step. An instance is shown in Figure \ref{instance}.

For the obtained reduced solution graph after \textbf{Steps 1-4}, an important question is whether any undiscovered conflict exist when different free nodes are assigned values? By the solution space expression of minimum vertex-cover, if one free node $i$ is applied the checking processes in \textbf{Steps 1-4}, it could be easily proved that node $i$ must be a free node in the whole solution space by the following process: set node $i$ as uncovered backbone and do the \emph{Freezing-Influence} propagation, if it passes \textbf{Step 3}, it can fix the states of some nearby nodes and no conflict exists, and then the solution space shrinks; choose any residual free node to fix its state and repeat the operation recursively, a complete minimum coverage solution can be obtained; similarly, set $i$ as covered backbone and another complete solution can also be obtained. Therefore, node $i$ is proved to have a free state in the solution space.

The above process provides a way to achieve the whole vertex-cover solution space of a KE graph, and it ensures that one reduced solution graph can represent all the possible solutions. This process costs time only on the maximum matching and \emph{Conflict-Checking}. Except the locating of maximum matching, the time complexity will cost no more than $O(n)$.


\section{Multi-Layer KE Subgraph Decomposition for Vertex-Cover}

\subsection{Definition of Multi-Layer KE Graph}

By the definition of the KE graph whose minimum vertex-cover number is equal to its maximum matching number, there are strong relations between the KE graph and its maximum matching. Here, we classify all the nodes into two separated classes $A_1$ and $B_1$ with $A_1\bigcup B_1=V$ and $A_1\bigcap B_1=\emptyset$ based on the maximum matching, where it is required that every two nodes on a same matching edge cannot be assigned into the same class and the unmatched nodes are all placed in class $B_1$ without loss of generality. Evidently there are exponential ways of arranging nodes into $A_1$ and $B_1$. The following result provides an alternative understanding of KE graphs.

\textbf{\emph{Theorem:}} For any KE graph $G$, there must exist some arrangements of $A_1^*$ and $B_1^*$ in which $B_1^*$ has no inner-class edges, i.e., all the edges in $G$ could only be inner-class edges in $A_1^*$ or the inter-class edges between $A_1^*$ and $B_1^*$, and any two matched nodes on the same matching edge could not belong to the same class. Correspondingly, a graph satisfying some arrangements of $A_1^*$ and $B_1^*$ above by maximum matching in which $B_1^*$ has no inner-class edges, must be a KE graph.

\textbf{\emph{Proof:}} $\Longrightarrow$ For a KE graph $G$, since its minimum vertex-cover number is equal to its maximum matching number, all the covered nodes must be the matched nodes and each matching edge has one and only one covered nodes. Choosing all the covered nodes in a minimum vertex-cover to form class $A_1^*$ and the rest assigned into $B_1^*$, all the nodes in $A_1^*$ should nake the coverage of all the edges in $G$, and it suggests that there is no inner-class edge in the class $B_1^*$. Furthermore, if there are different solutions of the minimum vertex-cover of graph $G$, there will be different arrangements of $A_1^*$ and $B_1^*$.

$\Longleftarrow$ If a graph $G$ has the arrangement of $A_1^*$ and $B_1^*$ where $B_1^*$ has no inner-class edges, all nodes in $A_1^*$ will lead to a vertex-cover of $G$. As the nodes number of $A_1^*$ is equal to the maximum matching number, it also determines the lower bound of the minimum coverage. Thus, the nodes in $A_1^*$ form a minimum vertex-cover, which results in $G$ a KE graph.  $\Box$

By the above theorem, a graph is a KE graph if and only if after deleting all the nodes in $A_1^*$ with related edges, the rest graph related with nodes in $B_1^*$ is a graph consisting of isolated nodes without any edge. In this paper, we also call a classical KE graph as 1-\emph{layer KE-graph}. If a graph $G$ is not a 1-\emph{layer KE-graph}, it implies that there exists no such $B_1^*$ that does not contain inner-class edges. In this case, our aim will be adjusted to find the \emph{proper separated classes} $A_1^*$ and $B_1^*$, in which the subgraph induced by $B_1^*$ has the lowest coverage number (smallest number of covered nodes) with respect to the minimum vertex-cover. Evidently, the KE graph can be viewed as a special case with the subgraph induced by $B_1^*$ having 0 coverage in this setting.

However, the core difficulty is to achieve the decomposition of $A_1^*$ and $B_1^*$, i.e., to find the subgraph induced by $B_1^*$ having the lowest coverage. In the proper classification $A_1^*$ and $B_1^*$, the subgraph induced by $A_1^*$ with the inter-class edges should occupy the coverage of vertex-cover the same as $\sharp A_1^*$ (nodes number in $A_1^*$) to make as many as covered nodes of $G$ belong to $A_1^*$: if some node in $A_1^*$ does not need to be covered, its matched node in $B_1^*$ can switch with it, namely switching their class labels. As a result, the decomposition of $A_1^*$ and $B_1^*$ is equivalent to obtain a subgraph with coverage number $\sharp C(G)-\sharp M(G)$, where $\sharp C(G)$ is the minimum coverage of $G$ and $\sharp M(G)$ is the maximum matching number of $G$. Since that minimum vertex-cover problem belongs to NP-complete problems and maximum matching belongs to P problems, the proper classification of $A_1^*$ and $B_1^*$ for general graphs is also a hard problem. Fortunately, it is easy and straightforward when the graph is a KE one, and the algorithm will be provided in next section.

\begin{figure}[htbp]
	\centering
	\includegraphics[scale=0.4]{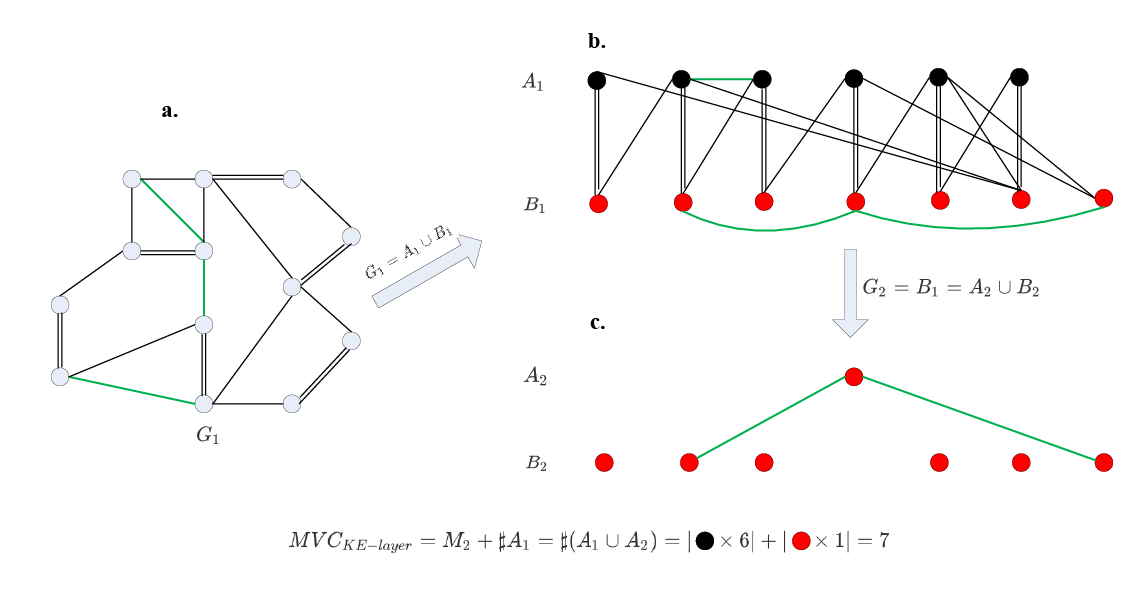}
	
	\caption{A 2-\emph{layer} KE graph example and its decomposing process. \textbf{a} is the original graph with one maximum matching marked by double-edge. \textbf{b} gives a classification in which nodes are assigned into $A_1$ and $B_1$: nodes on the same matched edge are not assigned into the same set and all unmatched nodes are placed into $B_1$ (only one node in this graph). There are two inner-class edges in $B_1$. In \textbf{c} the $G_2$ is the subgraph induced by nodes in $B_1$ and it could be decomposed into $A_2$ and $B_2$. There is not edge in $B_2$ in this case, which suggests that subgraph induced by $B_2$ is a KE graph and the vertex-cover of the original graph is consisted of $A_1$ and $A_2$. The cover number of the graph in \textbf{a} got by this strategy is 7, which exactly equals to the minimum cover number of this graph. }\label{decompose}
\end{figure}

Then, based on the above definition and analysis, if the subgraph $G_1$ induced by $B_1^*$ is not empty but a KE one, the graph $G$ is called a 2-\emph{layer KE-graph}, and this subgraph could find a \emph{proper separated classes} $A_2^*$ and $B_2^*$, in which $B_1^*=A_2^*\bigcup B_2^*$, $A_2^*$ and $B_2^*$ have the same meaning with $A_1^*$ and $B_1^*$ but restricted on $G_1$. When a graph is a 2-layer KE-graph, its minimum coverage is equal to the number of nodes in $A_1^*\bigcup A_2^*$. Similarly, we can define a $k$-\emph{layer KE-graph} and nodes in $G$ can be decomposed as $A_1^*\bigcup A_2^* \cdots \bigcup A_k^* \bigcup B_k^*$ with the class $B_k^*$ having no inner-class edges. In this way, the minimum coverage should be equal to $\sharp A_1^*\bigcup A_2^* \cdots \bigcup A_k^*$. The upper bound of the $k$-layer KE-graph satisfies $k\leq log_2{n}$, where $n$ is the node number of the whole graph and the upper bound will be achieved when the graph is a complete graph. A 2-\emph{layer} KE graph example and its decomposing is given in Figure \ref{decompose}.

\subsection{Algorithmic Framework for Measuring KE-layer}

In this section, a framework to find the KE-layer number for general graphs is proposed. Several approximation strategies will be presented to approach the KE-layer number and minimum vertex-cover of graphs. To measure the KE-layer number of a general graph, the maximum matching needs to be calculated at first and nodes are assigned into $A_1$ and $B_1$ as required above. Then the matched pairs of nodes will be switched to lower the coverage of $B_1$ to approach $B_1^*$. The framework is described in Algorithm \ref{measure_ke_layer}.

\begin{algorithm}[htbp]
	\caption{Framework for Measuring KE-layer}
	\begin{algorithmic}	         
	         \State  Let $l=0$, $A_0=\emptyset$, $B_0 = G$;
	         \While {$B_l$ is not KE graph}
	         \State $l=l+1$, $G_l =$ subgraph induced by $B_{l-1}$;
	         \State Calculate the maximum matching $M(G_l)$ of graph $G_l$;
	         \State Place each pair of matched nodes into $A_l$ and $B_l$;
	         \State Place every unmatched node into $B_l$;
	         \State $E=n$, $(E', A_l, B_l)=$ \emph{\textbf{Switch-Matched-Nodes}}($G_l$,$A_l$,$B_l$);
	         \While{$E>E'$}
	         \State Let $E=E'$, $(E', A_l, B_l)=$ \emph{\textbf{Switch-Matched-Nodes}}($G_l$,$A_l$,$B_l$);
	         \EndWhile
	         \EndWhile
	         \State $l=l+1$, $G_l =$ subgraph induced by $B_{l-1}$;
          	 \State Calculate the maximum matching $M_l$ of the subgraph induced by $B_l$;
	         \State \textbf{Return} KE-layer number $l$, vertex-cover number $\sharp M_l + \sum_{i=1}^{l-1}\sharp A_l$
	        \State ** \emph{\textbf{Switch-Matched-Nodes}}($G_l$,$A_l$,$B_l$) returns new $A_l$, $B_l$ and coverage of set $B_l$.     
	\end{algorithmic}
	\label{measure_ke_layer}
\end{algorithm}

%
%
%
In this framework, for a graph with KE-layer number $L$, $A_l$ with$1\leq l \leq L$ is peeled layer by layer until the remained subgraph $G_L$ induced by nodes in $B_L$ is a KE one. All nodes contained in $A_l$ and the minimum vertex-cover of $G_L$ will compose the vertex-cover of graph $G$. Since $G_L$ is a KE graph, it is fast to find its minimum vertex-cover. Thus, the vertex-cover number of graph $G$ with KE-layer $L$ could be calculated as:
\begin{equation}
	MVC_{KE-layer} = \sharp M_L + \sum_{i=1}^{L-1}\sharp A_l.
\end{equation}
During the whole process, the judgement of KE graphs on each $G_l$ could be implemented by Algorithm \ref{main_process} proposed above. An example of 2-\emph{layer} KE graph with detailed decomposing process to get its minimum vertex-cover is given in Figure \ref{decompose}

\subsection{Strategies for Switching Matched Nodes}

A key problem in this framework is how to place the matched pairs of nodes into $A_l$ and $B_l$ to lower the coverage of $B_l$ for each graph $G_l$, namely the \emph{Switch-Matched-Nodes} function above. Even with the constraint that every pair of matched nodes are placed into different sets, for a graph with maximum matching number $M$, there are $2^M$ possible arrangements, which makes it hard to find the optimal solution. This problem could be described as:
\begin{eqnarray}
	\textrm{$\mathop{\min}_{n\in A_l or B_l, \forall n\in V(G_l)}$\quad Coverage($G_l$)},\\
		s.t.\quad \textrm{$\forall$ matched edge $(i,j)$, $i\in A_l$, $j\in B_l$ or $j\in A_l$, $i\in B_l$.}
\end{eqnarray}

To solve it, several methods are proposed to optimize the solution. Two factors will determine the efficiency of this framework: the switching strategy of pairs of matched nodes and how to calculate the coverage of $B_l$.

\subsubsection{Switching Strategy}
A directed idea is to switch pairs of matched nodes in a greedy way. For each matched nodes pair, if switching them will lower the coverage of $G_l$, this switching should be kept; if the switching does not change or increase the coverage, this pair of nodes should be switched back. The detailed procedure is presented in Algorithm \ref{strategy1}, \textbf{Strategy 1}.

\begin{algorithm}[htbp]
	\caption{\emph{\textbf{Switch-Matched-Nodes}}(G, A, B), \textbf{Strategy 1}}
	\begin{algorithmic}
	\State $E =$Coverage of $B$;
	\For{every matched nodes pair $(i,j)$ with $i\in A$ and $j\in B$}
	\State Place $i$ in $B$ and $j$ in $A$, and calculate $E'=$ Coverage of $B$;
	\If $E'\geq E$)
	\State Place $i$ back in $A$ and $j$ back in $B$;
	\Else 
	\State $E=E'$;
	\EndIf
	\EndFor
	\State \textbf{Return} $E$, $A$, $B$
	\end{algorithmic}
	\label{strategy1}
\end{algorithm}
%

Obviously, this method is flawed: lowering the coverage in every step does not mean the global optimal solution. Another idea is to introduce some uncertainty and for each step, and here two pairs of matched nodes will be selected randomly and switched to lower the coverage. This process is presented in Algorithm \ref{strategy2}, \textbf{Strategy 2}.

\begin{algorithm}[htbp]
	\caption{\emph{\textbf{Switch-Matched-Nodes}}(G, A, B), \textbf{Strategy 2}}
	\begin{algorithmic}
	\State $temp = 0$;
	\While{$temp<\sharp M(G)$}
	\State $temp = temp +1$;
	\State Calculate $E =$Coverage of $B$;
	\State Randomly select two matched nodes pairs $(i_1,j_1)$, $(i_2,j_2)$, $i_1, i_2\in A$, $j_1,j_2 \in B$;
	\State Place $i_1$, $i_2$ in $B$ and $j_1$, $j_2$ in $A$, calculate $E'=$ Coverage of $B$;
	\If{$E'\geq E$}
	\State Place $i_1$, $i_2$ back in $A$ and $j_1$, $j_2$ back in $B$;
	\Else
	\State $E=E'$;
	\EndIf
	\EndWhile
	\State \textbf{Return} $E$, $A$, $B$
	\end{algorithmic}
	\label{strategy2}
\end{algorithm}

Ensuring that each switch lowers the coverage is not always a good strategy. Sometimes, to allow the increase of coverage at several steps may bring more possibility to the solution. So in the next strategy, when the switch does not lower the coverage, we keep the switch with a certain small probability and continue the process. This strategy is applied based on  \textbf{strategy 2} and is described in Algorithm \ref{strategy3}, \textbf{Strategy 3}.

\begin{algorithm}[htbp]
	\caption{\emph{\textbf{Switch-Matched-Nodes}}(G, A, B), \textbf{Strategy 3}}
	\begin{algorithmic}
	\State $temp = 0$;
	\While{$temp<\sharp M(G)$}
	\State $temp = temp +1$;
	\State Calculate $E =$Coverage of $B$;
	\State Randomly select two matched nodes pairs $(i_1,j_1)$, $(i_2,j_2)$, $i_1, i_2\in A$, $j_1,j_2 \in B$;
	\State Place $i_1$, $i_2$ in $B$ and $j_1$, $j_2$ in $A$, calculate $E'=$ Coverage of $B$;
	\State Generate a random number R;
	\If{$E'\geq E$ \textbf{and} $R\leq$ Threshold}
	\State Place $i_1$, $i_2$ back in $A$ and $j_1$, $j_2$ back in $B$;
	\Else
	\State $E=E'$;
	\EndIf
	\EndWhile
	\State \textbf{Return} $E$, $A$, $B$
	\end{algorithmic}
	\label{strategy3}
\end{algorithm}
%
%

\subsubsection{Coverage Estimation}
The coverage here refers to the minimum vertex-cover number of $G_l$, which is hard to get for a general graph. To estimate the coverage, a direct idea is to consider the edge number contained in $G_l$. For random graphs, when the average degree increases, with a high probability the vertex-cover number will increase. Another idea is to use the maximum matching number of $G_l$ to approximate its coverage. The maximum matching number of a graph determines its lower boundary of minimum vertex-cover number, which suggests that it is a good estimation for the coverage of the graph. When the maximum matching number is high, with a high probability, this graph owns a high minimum vertex-cover number. In the following sections, we use \emph{edge energy} and \emph{maximum matching energy} to denote these two coverage estimation.

\section{Experiments and Results}

In this section, a series of experiments will be conducted on random graphs. The switching strategies and vertex-cover estimating methods mentioned above will be implemented and compared. The KE-layer number and vertex-cover number of these graphs will be discussed.

\subsection{KE-Layer Number}

\begin{figure}[htbp]
	\centering
	\includegraphics[scale=0.43]{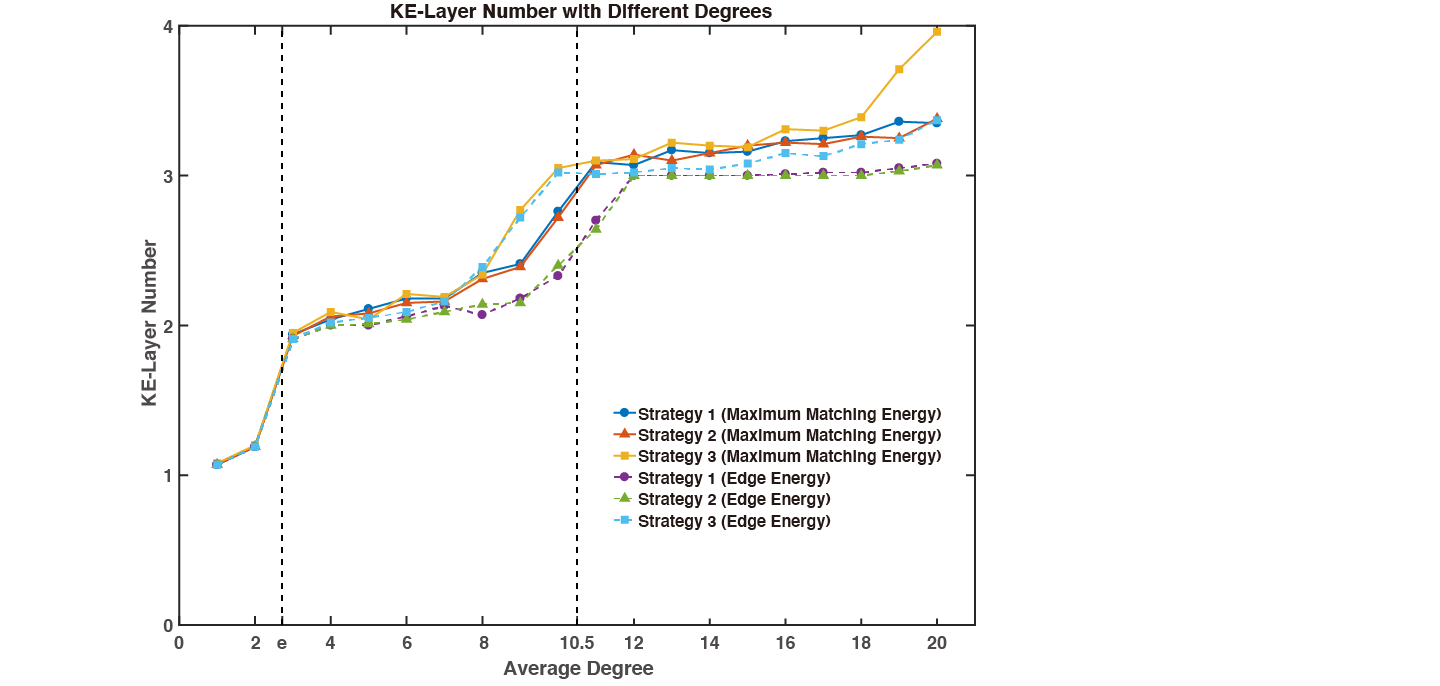}
	
	\caption{The KE-layer number by the proposed strategies on ER graphs with average degrees ranging from $1$ to $20$. For each average degree value, $100$ graphs with $1000$ nodes are generated. Results of \textbf{Strategy} \textbf{1}, \textbf{2}, \textbf{3} with both edge energy and maximum matching energy are drawn on the graph. The possible position phase transition points from KE-level 1 to KE-level 2 and from KE-level 2 to KE-level 3 are drawn on dotted line.}\label{KE_layer}
\end{figure}

The KE-layer numbers of ER random graphs with different average degrees are calculated and discussed in this subsection. Each graph contains $1,000$ nodes and for each average degree value, 100 graphs are generated and calculated. The average KE-layer numbers at each average degree are presented in Figure \ref{KE_layer}. As we could see, although different strategies perform different efficiency, there is a clear transition from 1-layer KE-graph phase to 2-layer KE-graph phase. The transition point is in the degree interval $(2,3)$. When average degree is higher than 12, the KE-layer number of almost all the graphs reaches 3. By the results in Figure \ref{KE_layer}, \textbf{Strategy} \textbf{1} and \textbf{2} with the edge energy have good performance and they will be analyzed in the following.

\begin{figure}[htbp]
	\includegraphics[scale=0.31]{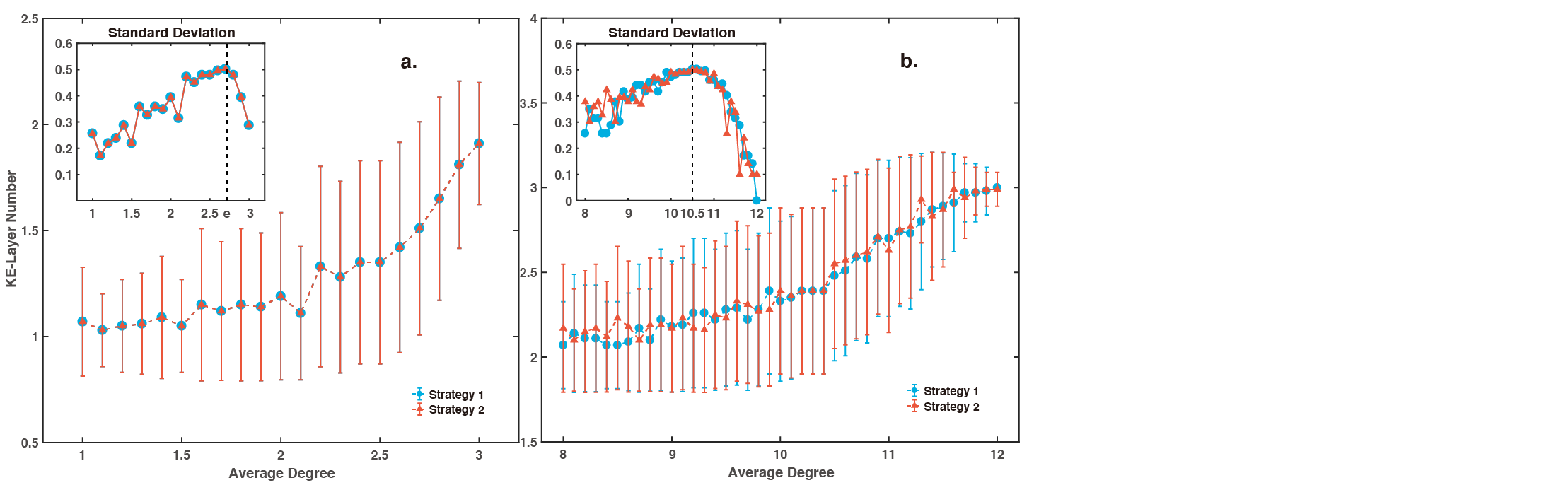}
	
	\caption{The KE-layer number errorbar by the proposed methods on ER graphs with various average degrees. For each average degree value, $100$ graphs with $1,000$ nodes are generated. Results of \textbf{Strategy} \textbf{1}, \textbf{2} with edge energy are drawn on the graph. The values of standard deviation are plotted in the inside figures. \textbf{a.} KE-layer numbers on ER graphs with average degrees ranging from $1$ to $3$. \textbf{b.} KE-layer numbers on ER graphs with average degrees ranging from $8$ to $12$.}\label{errorbar}
\end{figure}

The phase transition points will be estimated based on this fact: when the graphs are far from the phase transition points, most of them should have the same KE-layer number and the fluctuation (standard deviation) of their KE-layer numbers should be small, but the fluctuation should get increased when it approaches the transition points and reaches the highest at the transition point. For the location of the 1-layer KE-graph to 2-layer KE-graph transition point, by observing the standard deviation values in Figure \ref{errorbar} from \textbf{Strategy} \textbf{1} and \textbf{Strategy} \textbf{2} with edge energy, it could be found that the transition point is around $e$ (average degree), where the standard deviation reaches the highest and gets lower quickly after. For the location of the 2-layer KE-graph to 3-layer KE-graph transition point, a similar analysis could be conducted and the phase transition point is around $10.5$ (average degree). Also when the 3-layer KE-graph starts to get dominate (average degree larger than 12), the standard deviation is very close to zero, which suggests that almost all the graphs with this average degree are 3-layer KE-graphs. Certainly, there will be more $k$-layer KE-graphs and corresponding phase transitions with $k>3$ when the average degree gets increased over 20, which will not be discussed in this paper.

\subsection{Minimum Vertex-Cover Number}

\begin{figure}[htb]
	\centering
	\includegraphics[scale=0.42]{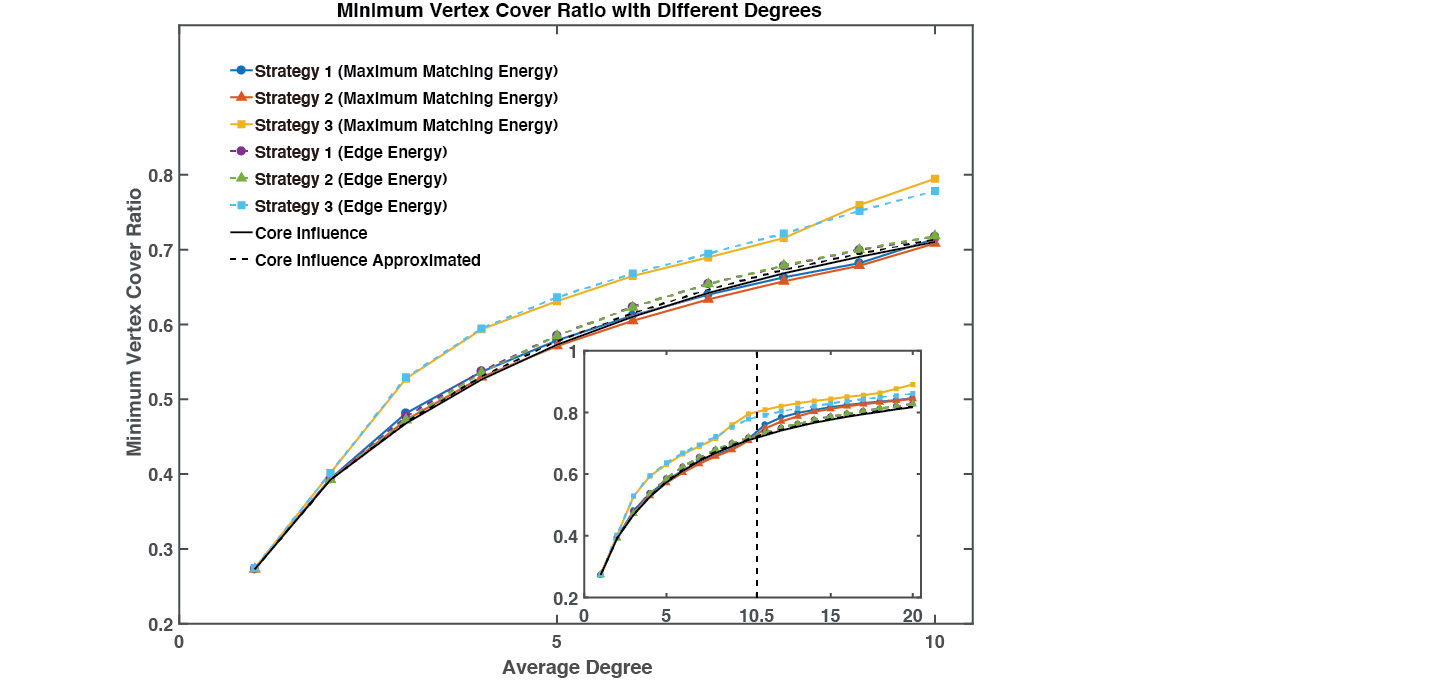}
	
	\caption{The vertex-cover ratio by the methods we proposed. The experiments are conducted on ER graphs with different average degrees. For each average degree value, $100$ graphs are generated and each graph contains $1,000$ nodes. Results of \textbf{Strategy} \textbf{1}, \textbf{2}, \textbf{3} with both edge energy and maximum matching energy, and core influence are drawn on the graph. The vertex-cover ratio with average degrees from $0$ to $20$ are plotted in the inside figure. The transition point $10.5$ is plotted, around which the vertex-cover ratio of \textbf{Strategy} \textbf{1} and \textbf{2} with edge energy surpass those with maximum matching energy.}\label{MVC}
\end{figure}

The process of calculating the KE-layer number could also approximate the minimum vertex-cover for the graphs. The experiments are also conducted on ER graphs, and in Figure \ref{MVC} the vertex-cover numbers of different strategies mentioned above are presented. 
As the Figure shown, before around average degree 10, the \textbf{Strategy} \textbf{2} with maximum matching energy works best compared to others. When the edge number contained in $G_l$ is used to estimate the coverage, results from \textbf{Strategy} \textbf{1} and \textbf{Strategy} \textbf{2} are quite similar, and perform the best when the average degree is higher than 10. An interesting phenomenon is that results of edge energy surpass the results of maximum matching energy at around degree 10. The reason is that when the average degree is higher than 10, \textbf{Strategy} \textbf{1} and \textbf{2} with edge energy could get lower KE-layer numbers, which lower the vertex-cover numbers at the same time. We also compare the results above with vertex-cover numbers from core influence \cite{feng2019core} in Figure \ref{MVC}. It could be observed that the KE-layer method performs better than core influence and could get lower minimum vertex-cover number with average degree lower than 10, and it performs almost same well with core influence when average degree is higher than 10.

To further study the efficiency of our KE-layer strategy, the vertex-cover numbers results from \textbf{Strategy} \textbf{2} with maximum matching energy are compared with the exact minimum vertex-cover numbers on small-scale random graphs. The differences between $MVC_{KE-layer}$ and real minimum vertex-cover numbers are calculated and the ratio of these differences against the total node numbers are presented in Table~\ref{small_graph}. As we could see, the KE-layer strategy performs well and when numbers of nodes increase the cover numbers stay stable. When the edge numbers increase with fixed node numbers, the gaps become larger, which is because of the increase of densely-connected clusters. Yet the KE-layer strategy still performs high efficiency and the vertex-cover numbers stay close to the real minimum vertex-cover numbers.

\begin{table}[htbp]
	\centering
	\caption{\label{small_graph} The percentages of differences between results of vertex-cover number from \textbf{Strategy 2} with maximum matching energy and exact minimum vertex-cover numbers against node numbers. The experiments are conducted on the ER graphs with $80$, $100$ and $120$ nodes with average degrees from $3$ to $7$ and every result is the average of $30$ graphs.}
	\begin{tabular}{cccccc}
		\hline
		\multirow{2}{*}{Node Numbers} & \multicolumn{5}{c}{Average Degrees}\\
		\cline{2-6} &3&4&5&6&7\\
		\hline
		$N=80$ & 0.88\% & 1.38\% & 1.83\% & 1.42\% & 1.92\%\\
		$N=100$ & 0.97\% & 1.77\% & 1.60\% & 1.63\% & 1.73\%\\
		$N=120$&0.47\% & 1.44\%& 1.61\% &1.81\% & 1.58\% \\
		\hline
	\end{tabular}
\end{table}

\section{Conclusion and Discussion}
In this paper, the graphical representation and hierarchical decomposition mechanism of Vertex-Cover is studied and based on this, the KE-layer structure for general graphs is proposed. A sufficient and necessary condition of determining KE graphs is given in the viewpoint of classifying the nodes into different classes based on the maximum matching, and an algorithm for verifying the KE graphs is provided, which reveals the structural feature and allows us to explore the hierarchical layer structure and the complexity of vertex-cover solution space evolution. Framework for calculating the KE-layer number for general graphs is discussed, and to solve the arranging and switching problem, several algorithmic strategies are compared including the switching strategies and coverage estimating methods. The phase transition points from 1-layer KE-graph to 2-layer KE-graph and from 2-layer KE-graph to 3-layer KE-graph are also located. At the same time, in this process, the vertex-cover number could be calculated and approximated, experiments are conducted on ER graphs to examine the efficiency, and the performance is related to the KE-layer complexity. This research provides a new perspective to approach the complexity of graphs and minimum vertex-cover problem.

A lot of research on this topic can be expected in future research. The switching methods could be further explored to improve the efficiency. Also, calculating the maximum matching for each step is very time-consuming, new convenient and fast coverage estimation measurement indexes deserve more exploration. At the same time, more other aspects related to this research could be expected. For example, some microscope structure, like motifs \cite{benson2016higher} and graphlets \cite{sarajlic2016graphlet}, may play crucial roles in the formation of KE graphs. Finding these structural components could improve the efficiency greatly. Also, relation between minimum vertex-cover and KE graphs could be further studied to find more frameworks to approach the complexity of NP problems.

\section{Acknowledgments}
This work was supported in part by the Research and Development Program of China (No.2018AAA0101100), the National Natural Science Foundation of China (No.62050132), and the Beijing Natural Science Foundation (China) (No.1192012, Z180005).

\section{Author Contribution Statement}
Xiangnan Feng conducted the experiments, wrote the main manuscript text and prepared Figure 2-4. Wei Wei conducted the analysis, wrote the main manuscript and prepared Figure 1. Jiannan Wang provided some advice for the experiments. Xue Liu provided some advice for the experiments and helped prepare the figure 2 and 4. Zhiming Zheng provided a lot of advice and suggestion for the experiments and manuscripts. All the authors have checked and approved the submitted version.

\bibliography{basename of .bib file}

\end{document}